\documentclass[sigconf]{acmart}

\usepackage{algorithm}
\usepackage[noend]{algpseudocode}
\usepackage{float}
\usepackage{eso-pic}
\def\BibTeX{{\rm B\kern-.05em{\sc i\kern-.025em b}\kern-.08emT\kern-.1667em\lower.7ex\hbox{E}\kern-.125emX}}
    
\setcopyright{rightsretained}
\acmConference[SIGIR 2019 eCom]{ACM SIGIR Workshop on eCommerce}{July 2019}{Paris, France} 
\acmYear{2019}
\copyrightyear{2019}
\makeatletter
\renewcommand\@formatdoi[1]{\ignorespaces}
\makeatother
\acmISBN{}

\AddToShipoutPicture{%
  \AtTextUpperLeft{%confidential
    \put(0,\LenToUnit{1.5cm}){\parbox{\textwidth}{\centering
       This paper was accepted to the ACM's Special Interest Group on Information Retrieval (SIGIR) workshops, 2019.}}
  }
}

\usepackage[font=small,labelfont=bf]{caption}
\begin{document}

%
% The "title" command has an optional parameter, allowing the author to define a "short title" to be used in page headers.
\title{Production Ranking Systems: A Review}

%
% The "author" command and its associated commands are used to define the authors and their affiliations.
% Of note is the shared affiliation of the first two authors, and the "authornote" and "authornotemark" commands
% used to denote shared contribution to the research.
\author{Murium Iqbal}
\affiliation{%
    \email{miqbal@overstock.com}
  \institution{Overstock.com}
}

\author{Nishan Subedi}
\affiliation{%
  \email{nsubedi@overstock.com}
  \institution{Overstock.com}
}

\author{Kamelia Aryafar}

\affiliation{%
    \email{karayafar@overstock.com}
  \institution{Overstock.com}
}

%
% By default, the full list of authors will be used in the page headers. Often, this list is too long, and will overlap
% other information printed in the page headers. This command allows the author to define a more concise list
% of authors' names for this purpose.
\renewcommand{\shortauthors}{Iqbal, Subedi, Aryafar}

%
% The abstract is a short summary of the work to be presented in the article.
\begin{abstract}
The problem of ranking is a multi-billion dollar problem. In this paper we present an overview of several production quality ranking systems. We show that due to conflicting goals of employing the most effective machine learning models and responding to users in real time, ranking systems have evolved into a system of systems, where each subsystem can be viewed as a component layer. We view these layers as being data processing, representation learning, candidate selection and online inference. Each layer employs different algorithms and tools, with every end-to-end ranking system spanning multiple architectures. Our goal is to familiarize the general audience with a working knowledge of ranking at scale, the tools and algorithms employed and the challenges introduced by adopting a layered approach.
\end{abstract}

%
% The code below is generated by the tool at http://dl.acm.org/ccs.cfm.
% Please copy and paste the code instead of the example below.
%


\maketitle

\section{Introduction}
% history of how we got here
The domain of ranking has its roots in the field of Information Retrieval (IR). Early automated IR systems, used in the 1950s were first applied to library indexing and employed statistics to retrieve documents from catalogs of thousands \cite{singhal2001modern, gleverdon1962report}. As the Internet has grown, an increasing number of industries rely on web and mobile platforms to reach end users. This has resulted in vastly larger catalogs of both public and private data. Ranking systems have emerged over time to extend the original IR systems to balance the goals of understanding user intent, scoring the relevance of an increasing number of items, and presenting users with results within fractions of a second. Organizations which use the Internet to interface with their users rely on ranking technologies to parse catalogs of millions or billions of items and surface the most relevant ones. These items range from music and movies available on streaming content services, to products for sale on e-commerce platforms, to web pages on the Internet cataloged by search engines, to advertisements for sponsored advertising and more. As such, ranking systems have become a core technology powering sales and user engagement. Users' interactions with the surfaced information is critical to the business of any such organization, and thus even a small improvement to these systems can yield significant growth for the business.

The need to increase user engagement has spurred an iterative experiment driven approach to improving ranking systems. The field has thus evolved into an intersection of research and application, with each system being built to simultaneously leverage complex machine learning (ML) methodologies and adhere to the constraints and tools required to support millions of users in real time. These methodologies select the most relevant items from catalogs of millions and present them in decreasing relevance to users. The need to accommodate experimentation with complex nonlinear models such as those based on deep learning, while still working within constraints, such as low latency, limited compute power and high parallelization, has driven ranking systems to evolve from a single system to a system of systems. These conflicting concerns are separated by isolating functionality within the systems. As such ranking systems are built with several layers of subsystems, including off line models which allow for flexibility of complex experimentation and online models which cater to live system constraints.

We view production ranking systems as having the following component layers: 

\begin{description}
\item[data processing] responsible for aggregating and featurizing raw data from various sources into training data for models
\item[offline representation learning] responsible for transforming raw data into embeddings or graph representations
\item[candidate selection] responsible for leveraging the learned representations to populate distributed databases with a selection of relevant candidates given a query
\item[ranking model] responsible for loading candidates from the distributed database or inverted index and ranking them in decreasing relevance given some context
\end{description}

Ranking systems are deployed to support recommendations, search, and sponsored advertising. In this work we examine production ranking systems as a general framework irrespective of their application. As such, in the context of this paper a query is generally the prompt to which the ranking system responds. This can be a text string used in search, a user or an item in recommendations, or a keyword in sponsored advertising. We use the term item to refer generally to any listing within a catalog, such as products for sale, advertisements, web pages, and more. We examine different approaches used in each of the layers of a ranking system across industries, but a convergence of methodology on any layer is still not apparent. Each individual application of ranking systems technologies requires it's own problem and data specific approaches to be developed. Instead of tabulating all approaches and caveats, which would be outside the scope of this work, we will examine the most popular general methodologies adopted for each layer of a ranking system.

The rest of the paper is organized as follows: Section \ref{sec:concerns} reviews how the conflicting goals of training models to rank items and serving models to support millions of users in real time has given rise to a system of systems. Section \ref{sec:processing} reviews ways to aggregate and normalize raw data into training instances for future layers. Section \ref{sec:representations} reviews various representations which are built to simplify the task of retrieval. Section \ref{sec:candidate_selection} reviews how learned representations can be leveraged by online ranking systems to select initial candidates. Section \ref{sec:online_models} reviews the live models used to infer item to query relevance and how they are served. Section \ref{sec:new_arch} reviews state of the architectures necessary to deploy ranking systems. Section \ref{sec:validation} reviews several ways to validate new components within a ranking system, and possible faults that can arise.

\begin{figure*}[t!]
\centering
        \includegraphics[scale=0.27,clip]{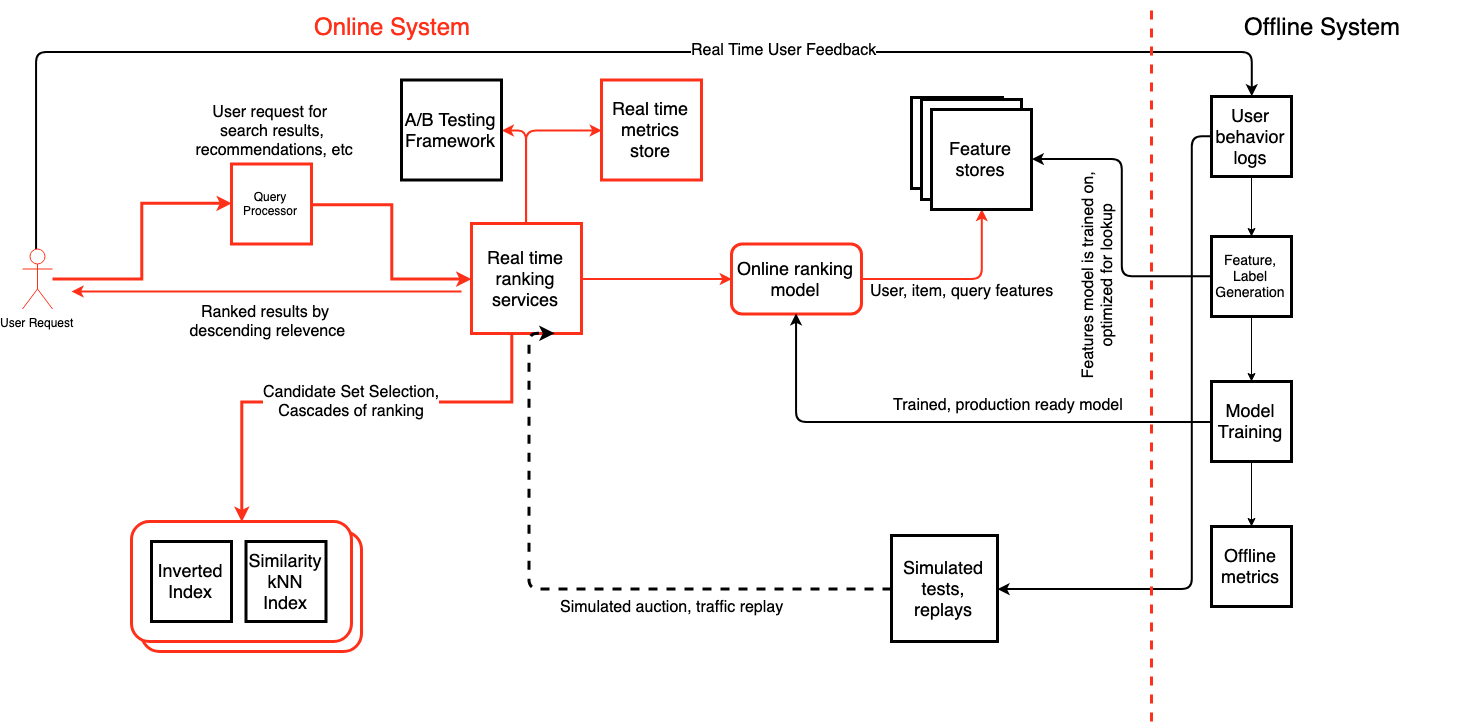}
\vspace{1mm}
\caption{General system architecture that covers the life cycle for a ranking request. The query can be items, user provided query strings or users in generalized ranking approach, and thus query processor can either act as a query re-writer, user personalization, or an item diversification layer.}
\label{fig:system}
\end{figure*}

\section{Separation of Concerns}
\label{sec:concerns}
The separation of concerns across the ranking system layers has allowed organizations to create models which can approximate ideal ranking, learn query intent, perform query re-writing, and diversify retrieved results. All of these applications are addressed by training offline models to capture relevant relationships within their representations. The necessary, often massive computations can be performed in batch, allowing for feature spaces which encode desired relationships within the geometry of the representation space. These relationships can then be directly leveraged by candidate selection or the online model. As the computationally heavy processes are captured in offline representation learning and candidate selection, online models can be built as less computationally complex models, e.g. linear or shallow models, allowing for low latency response time.

This separation allows for highly complex systems to be built and leveraged, but introduces failure points in the form of decreased interpretability of the ranking system behavior, difficulty in tuning, and increased difficulty of validation. This makes it hard to interpret experimental test results and improve upon previous iterations. Architectural decisions within the various layers of a ranking system also introduce corresponding assumptions into the overall system. Often these assumptions although necessary, can obfuscate biases and weaknesses.

Ranking systems have developed two architectures which have facilitated the separation of concerns. These are distributed technologies and one box models. As item catalogs have grown, technologies which require the housing of indices in memory are no longer feasible. As such, distributed databases which can house an index across a cluster of machines and coordinate retrieval from this index have become prevalent \cite{ghemawat2003google, shvachko2010hadoop, lakshman2010cassandra}. These distributed database technologies are able to house petabytes of data and run computations over their entirety. In parallel to the development of distributed computing technologies, ML models have grown in complexity, especially with the advent of deep learning. These models require specialty hardware to support their high computational complexity, such as CUDA enabled graphics processing units (GPUs) \cite{nvidia2010programming}. Often the computations necessary for these models are infeasible to translate to distributed system frameworks, and the volume of model parameter updates required to coordinate across the clusters is prohibitive to deploying these models on distributed systems. As such, one-box architectures are attractive. These architectures pull processed training data from distributed data stores and train complex models in memory on a single machine \cite{eksombatchai2018pixie, gupta2013wtf}.

An end to end ranking system employs both of these architectures, with distributed architectures providing a sink for raw data, embeddings, candidates and model predictions, and a source of training data and features for one-box models. This system of systems requires orchestration across frameworks, which can be handled via schedulers that are responsible for coordinating workflows for training and deployment of online models \cite{beauchemin2016airflow}. Each set of tools employed by the layers of the ranking model must be selected with care, as increasing the number of employed tools increases the complexity of the overall ranking system. This can cause some layers to be poorly configured, as practitioners are required to master many technologies. In some cases, separation of concerns can lead to isolation of practitioners who specialize on specific layers within the ranking system. This can cause further poor configurations, as practitioners may treat other layers of the system as black boxes, leading to layer specific optimizations which may give rise to suboptimal behavior across the entire system \cite{sculley2015hidden}.

We feel the separation of concerns is a useful tool, but only when employed with care. Research on productionized ranking systems tends to focus on just a single layer \cite{yan2018beyond, wu2018eenmf, wang2018billion, ni2018perceive}. Practitioners must be careful to examine the individual component layers as well as behavior across the entire system. Only the ability to interpret the system at both scales can yield an understanding of behavior and enable proper iteration to improve results. A diagram of the architectural components necessary to support a full ranking system is provided by in Figure \ref{fig:system}

\section{Data Processing}
\label{sec:processing}

\subsection{Datasets}
Increasing user interactions is the primary goal of ranking systems \cite{liu2009learning, covington2016deep}. As such, user interaction logs captured by the platform often serves to be the richest source of training data for the system. Some user interactions are ubiquitous, such as user clicks or user item ratings. Others are platform specific, such as purchases on e-commerce platforms, duration of viewing time on streaming content platforms, or likes on social media platforms. Data about the items themselves, such as title, category, associated text and cost, is referred to as side information. Side information is gathered by platforms either by user feedback, such as tagging on social media platforms, or by content providers, such as attribute labels provided by vendors on e-commerce platforms, or by the platform itself such as item categories.

Most data used by ranking systems are sparse high dimensionality vocabularies. This is especially true for user interaction data, where each item in a catalog can be seen as a word in the vocabulary with few user interactions \cite{zhou2018micro}. Side information contains a mix of sparse and dense data, such as sparse multi-hot encodings of text, or dense data of item price and size or item images. Prior to representation learning and model training, raw data must be normalized, featurized and formatted. To reduce noise and reduce computation time, large cardinality spaces can be trimmed via thresholding to drop highly sparse dimensions from the vocabulary. Out of vocabulary components can either be dropped or mapped to a default null representation \cite{wu2018eenmf}. Dense features which contain high variability or follow an exponential distribution can be normalized and/or smoothed prior to ingestion by models. 

\subsection{Data Aggregation And Normalization}
User interaction data can be aggregated by various methods, each building its own assumptions into the ranking system. Excepting reviews and ratings, user interaction data is often referred to as implicit feedback data. This is due to the fact that although platforms can present items to users and track specific user interactions, the relationship between those interactions is only implied. Negative signals, indicated by lack of user interaction, can also only be implied as it is impossible to know exactly what items were viewed by the user. Thus, a complete set of ranking labels for all items is impossible to capture \cite{joachims2002optimizing}. To reduce noise within this dataset, outliers, such as those associated with bot-like behavior or accidental clicks, are removed. This is generally done by thresholding and weighting user interactions by dwell time \cite{wang2018billion, grbovic2017search}. Selection of the date range over which to populate data can also affect models, as the volume and sparsity of data changes over the training window. Seasonal trends must also be accounted for when aggregating user interaction data. These aggregations, although necessary, and often specific to dataset and application, propagate assumptions through all layers of the ranking system and should be made with care \cite{sculley2015hidden}. 

Attributing user clicks to searches or recommendation carousels affects both model training and evaluation \cite{kannan2016path, geyik2014multi}. This is not limited to sponsored advertising and affects all ranking problems. For example, in gathering training data for search, do products associated with the search only include those clicked immediately after the search? Or should they include items clicked with subsequent searches assuming that these searches are refinements on the original search? Should two items clicked by the same user across days be considered related or only those clicked within the same hour be considered related? Adjusting these data aggregation layers and their underlying assumptions dictates which correlations are and are not captured within the dataset. Class labels on training instances per user click based relevancy can also change depending on the assumptions made \cite{wu2018eenmf}. These decisions can be seen as a form of data tuning which build the assumptions into the datasets used for both training, tuning and evaluating ranking systems. 

\subsection{Creating A Balanced Dataset}
The Learning To Rank (LTR) framework poses ranking as a supervised learning problem \cite{liu2009learning}. As such, both positive and negative samples need to be inferred from user interactions. Just as with positive samples, there are numerous ways to attribute negative samples, for example, if a user exited a video stream before finishing the entire video. There are far more possibilities for negative samples, represented in the lack of user interactions, than positive samples. Various methods for balancing the dataset are employed. One popular method is for each positive sample defined in the dataset only sampling one corresponding negative sample. Different mechanisms of negative item sampling are employed to select negative samples such that specific relationships are reflected in balanced pairs of negative samples and positive samples. This could include choosing negative items from categories that are far away from the item, excluding highly co-viewed or co-engaged products from the negatives list \cite{recommendations2018kangal}. Selecting negative interactions from only a window around a positive interaction can allow an assumption of user impression. 

Ranking systems have an intrinsic positional bias associated with them \cite{radlinski2006minimally}. Users click on higher presented results irrespective of query relevance, leading to false positive labels. Ignoring this bias and training on naively collected data can lead to models that simply fit the behavior of the existing global ranking function. The $FairPairs$ method modifies search results in a non-invasive manner that allows us to collect pairs of results that are unaffected by this presentation bias by randomizing the order of results between a small window of items during presentation. \cite{lynch2016images}.

\subsection{Discussion}
Although user interaction data is rich, incorporation of side information is necessary in any ranking system. This data is employed to combat popular item bias, and to address the cold-start problem \cite{lika2014facing}. This side information allows correlations learned from user interactions to be extended to items with few impressions based on relationships which exist in item descriptions, item category information, etc. Although proper processing of user interaction data is required for training high performance ranking models, we find effective ways to leverage side information affords the highest coverage and most diversity. Models which rely too heavily on user interaction data overfit to head queries and popular items.

Processing this data, which is often multiple gigabytes, must be done over a distributed computing platform, as it would be infeasible to fit the dataset in memory on a single machine. These processes are often done in batch, with user interaction data being processed into new training data at regular intervals. As the need for real time personalization and sponsored advertising increases, stream processing methodologies are emerging which allow data to be processed and featurized in near real time. This trend allows ranking systems to become increasingly responsive to user interactions and increase user engagement by modifying representations of users and query intent as users progress through a platform.

\section{Representation Learning}
\label{sec:representations}
Vector representations, learned over processed data, facilitate communication across the different layers of ranking systems. Representation learning leverages complex state of the art models, often relying on deep learning architectures. These models are highly platform specific, with architectures and solutions which yield large lift in one domain not necessarily providing effective representations on another platform. As such, much recent research on production ranking systems revolves around learned representations and much of a practitioners time is spent on this layer. These models are tuned to transform raw data into succinct expressive representations which can be used either for candidate set retrieval or reranking. These representations can be seen as mappings which project input data into low dimensional embedding spaces where distance is inversely related to relevance. Representation learning is performed offline, which allows this layer to leverage complex nonlinear ML architectures without the constraints of real-time systems. This allows the computational burden inherent in representation learning to be placed ahead of the live ranking layer. These representations are often learned from multiple modalities, incorporating both user interaction data and side information. We examine several methods of learning representations from these modalities both jointly and separately.

\subsection{Shallow Embeddings}
The simplest architectures to form representations employ shallow architectures such as word2vec over a combination of vectorized interaction data and side information. Here a "word" within the vocabulary can be categorical user actions, a discretized continuous feature, or English words \cite{grbovic2016scalable}. User action data can be taken strictly as words, or as sequences of actions, in which case embeddings can be built from skip-grams of user action sequences \cite{wu2018eenmf, grbovic2016scalable}. Multi-hot user action vectors can also be weighted by dwell time \cite{grbovic2016scalable}. This allows us to learn query and item representations unsupervised from user engagement data. To handle cold start, out of vocabulary items can be taken as linear combinations of embeddings of associated in vocabulary items, or corresponding content data \cite{grbovic2016scalable, recommendations2018kangal, wu2018eenmf}.

\subsection{Multi Modal Representations}
User interaction, text based side information and image based side information are each distinct datasets from various sources that follow distinct distributions. They are, however, related in that they express information about the same items. To simultaneously leverage all datasets and learn a single representation ranking systems employ multimodal learning \cite{ngiam2011multimodal}. The simplest approach to this end is concatenation of raw vectors to generate a single input to a model which learns a representation \cite{ni2018perceive}. Although simple, this method fails to take advantage of the separate structures within each modality. Another approach is to train separate embedding layers for each modality and use these as input to another model which learns a combined embedding. This methodology allows separate architectures to be employed for each modality \cite{wang2018billion}, but adds complexity in that the individual architectures have no obvious method of validation. Furthermore if one modality of data is not present for an item, this could cause unexpected results in the final representation.

Extensions to this method train separate embeddings for each modality but with a cost function used across the separate models to force them to map to the same space \cite{wang2018billion}. Items can then be taken as a weighted sum of the embeddings of their associated data points. This method allows for items with incomplete side information. Mapping modalities to the same embedding space can also be performed in a the methodology of $search2vec$ \cite{grbovic2016scalable}, in which vectors built from side information are first initialized to the corresponding user action vectors. Side information is then sampled to form n-grams which are subsequently embedded into the same space as the user action vectors, and only those within a minimum cosine similarity to the user action vector are kept. Embeddings can be trained jointly over modalities, by employing siamese networks \cite{kang2017visually}, or by allowing certain subgraphs within a network share weights \cite{wu2018eenmf}. In these architectures side information is used as features to learn supervised embeddings with the user interactions providing the relevance to encode within the space. This methodology introduces training and architectural complexities but provides powerful representations. 

\subsection{Multi Task Learning}
Multi-task learning aims to train robust representations by jointly training representations for multiple applications \cite{ni2018perceive, wu2018eenmf}. Each task can be viewed as a regularization to the other tasks. This can yield powerful generalizable representations, but requires tasks to be related. Multi-task learning can yield unstable results when poorly configured.

\subsection{Graph Representations}
User action data can also be represented as graphs instead of vector embeddings. Here nodes are individual items and edges are user interactions. Graphs are initialized with raw user interaction data and side information. The final structure of the graph is learned by training models to prune edges within the graph via logistic regression, gradient boosted decision trees or multi-layer perceptrons \cite{yan2018beyond, eksombatchai2018pixie}. Hierarchical graph structures can also be built, where each level of the graph represents a type of node. For example, in sponsored search setting, the first tier can represent a query signal, the second tier can represent keywords, and the third tier can represent ads. These representations can be used directly with graph based candidate selection methods without the use of embeddings as described in Section \ref{sec:candidate_selection}. Once in graphical form, random walks can be employed to transform graph representations into sample data points. These samples can then be used as input to an embedding model similar to the raw sequences of user clicks. Embeddings built off of these samples purportedly capture higher-order item similarities than direct sampling of user interaction sequences \cite{wang2018billion, eksombatchai2018pixie}.

\subsection{Discussion}
As state of the art work in deep learning continues to produce more effective representation learning techniques, we find the best approach is to employ embeddings which can surface relationships within the underlying data. Validation of this layer is difficult, and not often discussed in the literature. Instead this layer is often validated in conjunction with the subsequent layers. This can cause improvements within this layer to be hidden by poor tuning in these subsequent layers. As such effective use of representation learning requires development of clear validation of this layer in isolation to the ranking system.

\begin{figure*}[t!]
\centering
        \includegraphics[scale=0.5,clip]{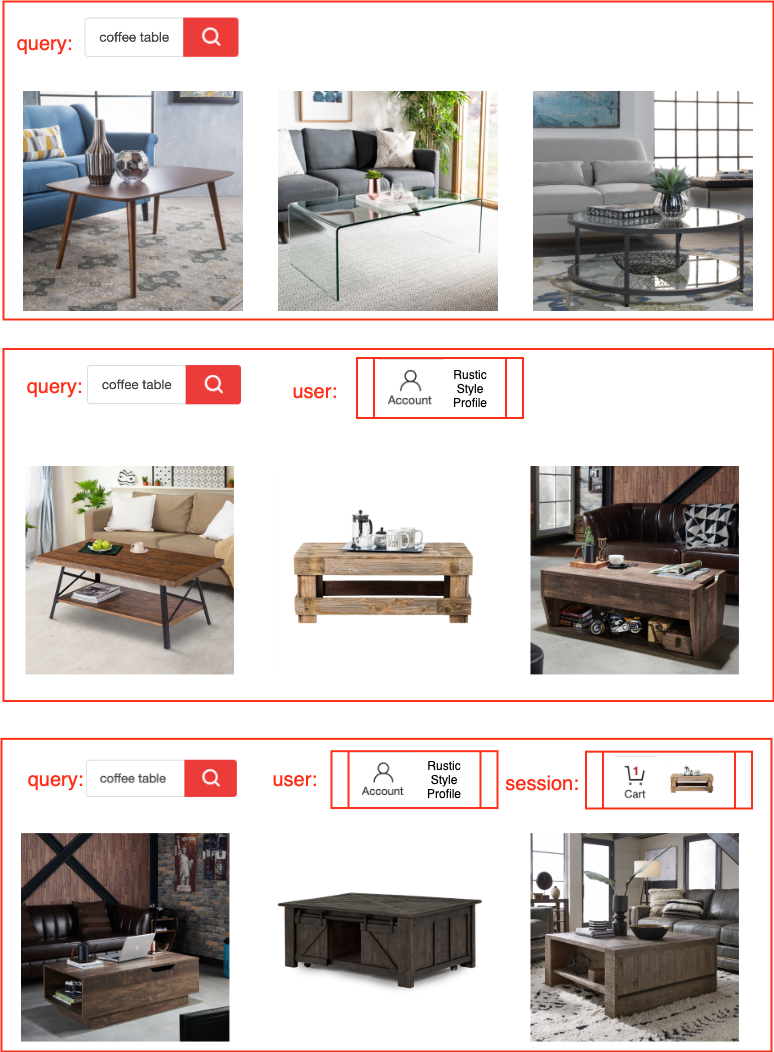}
\vspace{1mm}
\caption{This figure highlights how ranking systems enable real time interaction with users. Users interact with the platform and the system processes their interactions to update their query results. The first row illustrates a global result. The second row shows results for the same query when the user signs in, and provides context based on their past history. The third row presents results for the user for the same query once a product has been added to the cart. This updates the context to the ranking model by providing an additional item, thus narrowing down to the user's information need. At each stage the system is dynamically ranking candidates after interpreting new information about the user to better infer their intent and each items relavance to the intent.}
\label{fig:realtime}
\end{figure*}

\section{Candidate Selection}
\label{sec:candidate_selection}

In ranking systems, candidate selection functions over the learned representations output from the offline models and populates databases which can be read from rapidly by online models. It should be noted that representations used for candidate selection can be distinct and separate from representations used as features to the online model. Representations used by candidate selection support projecting queries into a shared representation space with items that encapsulates similarity, query intent, and support personalization. The goal of candidate selection algorithms is to use the representations to populate a distributed database with a relatively small set of candidates for each query. \cite{grbovic2016scalable}. 

After offline systems build embedding spaces in which spatial relationships encode relevance, candidates are selected by nearest neighbor searches. A query is represented in the embedding space, and all items within the catalog closest to it, given some distance metric, are selected as candidates. As directly computing exact $k-NN$ from catalogs of millions is prohibitively computationally expensive, various methods for approximate nearest neighbor (ANN) searches are employed. All of these methods populate an optimized lookup index. Approximate nearest neighbor methods can be broadly broken into three categories, hash based, tree based and graph based. 

\subsection{Approximate Nearest Neighbors}
\paragraph{Hash based}
Hashing based approaches, such as locality sensitive hashing or FALCONN \cite{razenshteyn2018falconn}, are simple models which can be scaled using distributed frameworks as each item can be hashed independent of others within the catalog. These methods compress high dimensional data, via hashes, and assume similar items' hashes will result in collisions \cite{aumuller2019ann}. This form of ANN has drawbacks in that in high dimensions false positives and false negatives can appear, as the data is highly sparse and the randomness inherent in the selected hashing function can incorrectly cluster items due to the curse of dimensionality \cite{verleysen2005curse}. 

\paragraph{Tree based}
Tree based methods are frequently built as in-memory models with many open source implementations available \cite{muja2015fast, bernhardsson2018annoy, naidan2015non}. Trees are built to be balanced by applying splits along different dimensions of input data. Nearest neighbors search is then performed by traversing the tree starting at the query node and finding nodes within minimum traversals. This method works well on low dimensional data, but at higher dimensions, performance degrades as tree based approaches are complex, often relying on several trees to obtain high performance and traversal of several trees is time consuming \cite{muja2014scalable}.

\paragraph{Graph based}
Unlike hashing and tree based approaches, graph based approaches function over graph representations of raw data, instead of embeddings. Graphs can be initialized with all incidences of user interaction indicating an edge, which populates a highly dense graph. The graph is then pruned, with models trained to learn relevance \cite{yan2018beyond, eksombatchai2018pixie}. Neighbors are discovered by employing navigable small world (NSW) or walk based proximity algorithms \cite{malkov2018efficient}.

\subsection{WAND}
Weighted And (WAND) method is a candidate selection algorithm adopted by some production ranking systems \cite{borisyuk2016casmos}. This method matches queries directly to items using raw features. Items are scored to relevance with queries by a weighted average of all features which they have in common with the query. The top scored items are taken as candidates. This method requires a weight matrix to be learned via constrained feature selection, but provides low latency response with minimal model complexity.

\subsection{Serving Candidates}
The databases used to serve the results from candidate selection are distributed frameworks to allow for rapid responses. These are updated in batch, with new candidates being populated on regular schedules. Search and advertising systems leverage distributed inverted index technologies such as Solr and Elasticsearch \cite{grainger2014solr, gormley2015elasticsearch}. Other similar architectures leverage distributed key value stores which prioritize high availability, such as Cassandra or Redis \cite{lakshman2010cassandra, carlson2013redis}. These systems are built to be highly fault tolerant and scalable to support growing numbers of users and items.

\subsection{Discussion}
Concerns with this layer of ranking and form of separation is propagation of error from new embedding experiments leaking into later layers due to proper lack of tuning of this layer. We find that hashing based methodologies are likely the most relevant to ranking, as much work is spent on representation learning. Graph based approaches are weak in that they require massive datasets, and only the largest organizations can afford to employ them. Graph based approaches have the further complication of lacking an obvious approach to addressing the cold start problem, as graph based representations cannot directly represent items which do not already exist within the graph.

\section{Ranking Models}
\label{sec:online_models}
Online ranking models return candidates in descending order of relevance to a query. Shallow models are used for real time services, making use of offline learned representations as input features \cite{ni2018perceive}. This allows online models to optimize for latency, limited storage and limited compute available to real time services. Figure \ref{fig:realtime} depicts how the online ranking inference models allow platforms to better interact with and engage users. There are three distinct ways to formulate the ranking problem to train online models; these are pointwise, pairwise and listwise approaches \cite{liu2009learning}.

\subsection{Pointwise Approach}
In pointwise approaches a model is trained to individually score the relevance of each candidate to the query. The problem is formulated as a binary classification, with a positive class indicating a user interaction such as a click or purchase, and the negative class indicating a lack of interaction. Training instances are taken from logs of user interactions. The model provides a score for each candidate reflecting its probability of eliciting an interaction given the query and any context of the user and items captured in the provided features. Candidates can then be ranked in descending order of likelihood for an interaction. This approach affords the use of shallow models, such as logistic regression. This methodology suffers from a lack of context about other candidates, as each candidate is scored individually. This causes an assumption that the output space of the candidates is a multi-variate Bernoulli, where each candidate's score is independent of each other. This is a poor assumption, as users view several items at once, and choose from among them. As such an item's probability of being clicked is affected by its neighbors. This approach is detailed in Algorithm \ref{alg:pointwise}.

\begin{algorithm}[b!]
\caption{Pointwise Approach (Users, Queries, Clicks)}
\begin{algorithmic}[1]
\State Nonclicks $\gets$ sample\_nonclick(c) $\forall$ c $\in$ Clicks
\State Labels, Interactions $\gets$ stack([0, Nonclicks], [1, Clicks])
\State Features = featurize(u, q, i) $\forall$ tuple(u,q,i) $\in$ (Users, Queries, Interactions)
\State P $\gets$ sigmoid($W^T$(Features) + $\vec{b}$)
\State C $\gets$ Labels $\times$ log(P)
\end{algorithmic}
\label{alg:pointwise}
\end{algorithm}

\subsection{Pairwise Approach}
In pairwise approaches a binary classifier is trained to score a pair of candidates simultaneously. The positive class indicates that the first candidate is more likely to be interacted with than the second, and the negative class indicates the opposite. Training data is initialized with all positive classes and balanced by randomly swapping the order of items within a pair with a fifty percent chance. This is the most popular approach as it allows for the model to consider relationships between candidates, rather than scoring them independently. This approach requires an additional sorting to be performed based on the scoring which causes further computation overhead. Pairwise approaches allow for regression models to be used, such as logistic regression and gradient boosted decision trees. Although gradient boosted trees provide better results, the logistic regression still provides the lowest latency response. Popular loss functions used to train these models are binary cross entropy and $\lambda$ loss function utilized by LambdaRank \cite{burges2010ranknet}. This approach is detailed in Algorithm \ref{alg:pairwise}.

\begin{algorithm}[b!]
\caption{Pairwise Approach (Users, Queries, Clicks)}
\begin{algorithmic}[1]
\State Nonclicks $\gets$ sample\_nonclick(c) $\forall$ c $\in$ Clicks
\State S $\gets$ s $\sim$ Bernoulli $\forall$ c $\in$ Clicks
\State Labels, Interactions $\gets$ stack(swap\_or\_not(s, c, n)) $\forall$ tuple(s, c, n) $\in$ (S, Clicks, Nonclicks)
\State Features = featurize(u, q, i) $\forall$ tuple(u,q,i) $\in$ (Users, Queries, Interactions)
\State P $\gets$ sigmoid($W^T$(Features) + $\vec{b}$)
\State C $\gets$ Labels $\times$ log(P)
\end{algorithmic}
\label{alg:pairwise}
\end{algorithm}

\subsection{Listwise Approach}
Listwise approaches require models to be trained over an entire list of items simultaneously. Formulation of a loss function for such a model is difficult, as the true ranking of an entire list is not possible to populate. One method extends pointwise approaches by assuming a multinomial instead of a multivariate bernoulli. This methodology forces candidates to compete with one another for limited probability mass. Other methods try to directly maximize NDCG as their objective \cite{taylor2008softrank, xu2007adarank}. This approach is detailed in Algorithm \ref{alg:listwise}.

\begin{algorithm}[b!]
\caption{Listwise Approach (Users, Queries, Clicks)}
\begin{algorithmic}[1]
\State Nonclicks $\gets$ sample\_nonclick(c) $\forall$ c $\in$ Clicks
\State Labels, Interactions $\gets$ stack([0, Nonclicks], [1, Clicks])
\State Features = featurize(u, q, i) $\forall$ tuple(u,q,i) $\in$ (Users, Queries, Interactions)
\State P $\gets$ softmax(sigmoid($W^T$(Features) + $\vec{b}$))
\State C $\gets$ Labels $\times$ log(P)
\end{algorithmic}
\label{alg:listwise}
\end{algorithm}

\subsection{Discussion}
The most common approaches for real time models are shallow binary classifications, especially those relying on pairwise approaches, which can rank an entire set of candidates against one another while still posing the problem as a binary classification\cite{ni2018perceive}. These approaches are best suited for machine learning, as binary classifiers are a well studied set of models. We view both pointwise and pairwise approaches as being the most optimal solutions. The challenge with these models is addressed by careful feature engineering in data processing and representation learning layers. Listwise approaches seem ill-posed in comparison, as true ranking data is impossible to obtain with implicit feedback and the models suffer from complexities of trying to learn relevancy of multiple classes at once.

\section{Serving Complex Models}
Different architectures are used to deploy various types of ranking systems. Each architecture has specific purposes and supports various layers of the overall ranking system.

\label{sec:new_arch}
\subsection{Distributed Architecture}
Distributed databases employ a cluster of machines and coordinate data storage and computations across the cluster. This architecture can be leveraged for data processing layers of ranking systems to aggregate, normalize, and featurize training data \cite{gauci2018horizon, park2018deep}. Recent work has also employed this architecture for representation learning via proprietary distributed computing technologies, such as those used for distributed computation of embeddings \cite{grbovic2016scalable, gauci2018horizon} or via open source libraries such as those available in Spark \cite{zaharia2010spark}. Several forms of candidate selection, those based on hashing, can also be performed on distributed databases. These architectures are used to serve learned representations and results from candidate selection to online models. Distributed databases further provide the ability to simultaneously write billions of records, affording the ability to capture user interaction data from millions or billions of users simultaneously. Distributed databases also afford high scalability, as new machines can be added to a cluster ad hoc to support increases in volume of data. These systems have their own limitations though. One such limitation is described by the CAP theorem which maintains that a distributed data store can not simultaneously provide consistency, availability and partition tolerance \cite{gilbert2002brewer}. Thus each distributed framework trades off one of these goals for the others, some prioritizing rapid capturing of user interaction with others prioritizing high availability to support real time response. This may require a single ranking system to employ several different distributed datastores, one to capture and process data, another to make data highly available to online models. Distributed data stores also require computations to be written within specific programming paradigms, such as MapReduce, which do not easily represent deep learning computations or graph based computations \cite{gupta2013wtf}. Furthermore, model training and prediction on distributed data stores require transference of model weights across entire clusters, which is infeasible for complex deep learning models with many parameters.

\subsection{One-box Architecture}
One-box architectures allow for the use of complex modeling techniques to be employed in ranking systems. Here a single machine loads training data from a distributed database and trains a model in memory \cite{eksombatchai2018pixie, gupta2013wtf}. These architectures are used for representation learning via deep learning models or graph based approaches. Candidate selection can also performed on one-box architectures via tree or graph based methods, which are described further in Section \ref{sec:candidate_selection}. Online models are often trained via one-box solutions, and can be served in parallel to scale to support requests from high volumes of users, such as millions or billions \cite{bernstein2014containers}. This architecture affords ML practitioners the freedom to use different tools, without the constraints imposed by distributed architectures. Recent development of containerized solutions allows one-box architectures to simultaneously support a number of tools and solutions by isolating system dependencies \cite{boettiger2015introduction}. Practitioners are thus free to employ complex models without the need to work within system specific paradigms. One-box architectures are limited to in-memory computations and cannot fully leverage the entire dataset and thus are still reliant on distributed datastores to pre-compute data.

\subsection{Serving Deep Learning Models}
As research on deep learning has progressed, a push to use these complex models in real time ranking applications has been made. This is a divergence from prior architectures which employ shallow models for real time inference. Several methods have been employed to enable deep learning models to be served and respond in real time. Each has with its own assumptions and trade-offs.

Deep learning training can be scaled by distributing training over multiple GPUs by leveraging recent open source tools, such as Horizon \cite{gauci2018horizon}. These tools provide support for training on CPU, GPU and multi-GPU in one box architectures and can provide the ability to conduct training on many GPUs distributed over numerous machines. This framework requires GPU servers, which can be cost prohibitive to purchase and maintain. 

Other platforms allow for distributed embeddings by adopting a parameter server (PS) paradigm which employs a cluster of servers. Model updates are communicated across the cluster through a central parameter server. These systems are designed with the following constraints to allow for fast training: Column-wise partitioning of feature vectors among PS shards; No transmission of word vectors across the network; PS shard-side negative sampling and computation of partial vector dot products \cite{dean2012large}. Here a shard is a subset of the entire dataset which is acted upon independently by one server in the cluster. These systems are complex to maintain and support as updates are made asynchronously to the model over each shard.

Embedding layers within deep learning models are fully connected, requiring many parameters. These computations are incredibly expensive. To allow for real time embedding of raw signals, model compression via quantization of model parameters is employed \cite{park2018deep}. For example, in some layers floating point precision of weights is reduced to 8 and 16 bits. This comes with a reduction in precision, but allows for a smaller memory foot-print. In this approach representation learning must be done such that models and their hardware are co-designed. The complexity of training such a model is much higher. Different types of quantization are performed given the acceptable drop in precision at each layer. Each layer is individually optimized, as well as the entire graph as a whole. Reduction of model parameters can be performed by model architecture decisions as well. For example, in the case of sequential models, GRUs are chosen to learn representations instead of LSTMs as they require fewer parameters \cite{wu2018eenmf}.

To allow model training to be completed in a timely manner, models with many parameters trained over large datasets can be trained incrementally, with a one-time training occurring infrequently and weight incremental updates being calculated daily \cite{ni2018perceive}.

\subsection{Discussion}
Current design paradigms rely heavily on the coordination of tasks across both distributed and one-box architectures. Data processing and candidate selection are performed distributed, while representation learning and online models are handled via one-box architectures. Tasks are coordinated via schedulers \cite{beauchemin2016airflow} and one box architectures are containerized and deployed in parallel to support requests from many users \cite{bernstein2014containers}.

We find that recent trends aim to allow service of complex models directly by making complex models more efficient \cite{park2018deep, yan2018beyond, wu2018eenmf, dean2012large}. This methodology has the potential to reduce the layers and separation of concerns within ranking systems. This would be a powerful improvement as it would reduce the complexity of the overall model, allowing for a unified approach. Improvements to streaming data processing technologies could further support these developments as complex computations can be run over distributed data caches in near real time to transform data into input features for deep learning models which are served live.  Although promising, this allows for deep learning models to be deployed in real time, but the work lacks a generalized approach, lacks open source support and thus lacks wide spread adoption.

\section{Validating A System of Systems}
\label{sec:validation}
We have thus far shown that a single ranking system is composed of several layers, each with its own complexity. The need for experimentation to increase user engagement requires both offline validation and online test results. Validating a ranking system, which spans several frameworks is not straightforward. Each individual layer should be validated in isolation in addition to the whole, but often in ranking systems, the reported results are only those of the candidate selection or the final ranking model. Such complex systems thus lend themselves to a change one thing change everything (CACE) data dependency and system entanglement \cite{sculley2015hidden}. This makes metrics unreliable, obfuscating errors and making it difficult to prove improvements to the overall system. Despite these reservations, both offline and online testing is performed primarily after candidate selection and after ranking. Depending on application the system and its component layers are generally tested for improvement on click-through-rates, purchases, dwell times and advertisement engagement to name a few. Selection of the desired metric to optimize for must be done with user behavior in mind, but often suffers from biases and is also affected by functionality outside of the ranking system itself, such as user interfaces.

\subsection{Increasing experiment bandwidth}
Speed of experimentation is hindered by the bandwidth for online testing, as there is a finite amount of traffic that lends itself to each particular test. One approach to improve throughput of testing involves early detection of poor or invalid experiments. This afford greater throughput of experiments as poorly performing tests are detected and terminated early. To allow for this standard metrics are populated frequently and made consistent across all related experiments. A multi-layer experiment architecture can also be employed where experiments are grouped into statistically independent layers. Each user is then simultaneously used as a data point for multiple tests, one from each layer, allowing multiple tests to be run simultaneously \cite{tang2010overlapping}. Experiment duration can also be decreased by employing variance reduction techniques, which separate users within the test group into two strata: those with prior purchase behavior and those without. For those with prior purchase behavior, this past data is used as a control covariate for additional variance reduction. This has been shown to reduce the duration of experimentation by half while maintaining equivalent confidence \cite{deng2013improving}.

\subsection{Offline evaluation}
Even with such methods, bandwidth for tests is limited. As such new model experiments must prove a significant improvement on offline validation to be selected for a live A/B test. Common offline metrics used are area under the curve (AUC) for the receiver operating characteristic (ROC) and normalized discounted cumulative gain (NDCG) as these correlate well with expected click through rate \cite{wu2018eenmf, yan2018beyond, grbovic2016scalable, ni2018perceive}. Simulated experiments can also be employed to estimate performance of models prior to A/B tests. These simulations must be calibrated to avoid incorporating bias leading to poor estimates of expected click through rate \cite{bai2018practical}.

\subsection{Discussion}
Focus on ranking metrics for the overall system is necessary, but we propose that each layer requires its own independent metrics as well to avoid obfuscating errors and biases. Data processing layers should document assumptions with metrics dashboards, and gauge distributions of data as well as any underlying shifts within these distributions over time. Validation on learned representations is not documented in most production ranking system architectures, instead they are only measured in their improvement of applications for modeling and candidate selection. Requiring each component to have independent functionality tests as well as tests of the entire system can more clearly surface errors \cite{gauci2018horizon, sculley2015hidden}.

\section{Conclusion}
We examine production ranking systems and find that a layered approach is adopted in every case. This is necessary to offset the computational cost of leveraging the most effective machine learning models, which are unable to produce real time inference for users. This layered approach causes ranking systems to be composed of a system of systems, each layer employing different algorithms over different architectures. This approach allows for rapid experimentation both within and across layers and allows practitioners to employ state of the art modeling techniques while still adhering to real time service constraints such as low latency and limited available memory. However, this same layered approach causes ranking systems to be incredibly complex, with each layer introducing its own assumptions and requiring its own tuning. This can obfuscate errors and makes it difficult to measure iterative successes. As ranking systems develop new methods to facilitate the direct serving of more complex systems, reliance on this layered approach could be reduced.

\bibliographystyle{ACM-Reference-Format}
\bibliography{references}

\end{document}